\documentclass[fleqn,usenatbib]{mnras}

\usepackage{newtxtext,newtxmath}
\usepackage[dvipsnames]{xcolor}
\usepackage{threeparttable}
\usepackage{multirow}
\usepackage{booktabs}
\usepackage{ulem}
\usepackage{makecell}
\usepackage{mathrsfs}
\usepackage{subcaption}

\usepackage[T1]{fontenc}

\DeclareRobustCommand{\VAN}[3]{#2}
\let\VANthebibliography\thebibliography
\def\thebibliography{\DeclareRobustCommand{\VAN}[3]{##3}\VANthebibliography}

\usepackage{graphicx}	
\usepackage{subcaption}
\usepackage{amsmath}	
\usepackage{amssymb}	
\usepackage{cellspace}
\usepackage{makecell}
\usepackage{longtable}
\title[]{CURLING - \uppercase\expandafter{\romannumeral2}. Improvement on the $H_{0}$ Inference from Pixelized Cluster Strong Lens Modeling}

\author[Y.-S. Xie et al.]{
Yushan Xie$^{1, 2}$,
Huanyuan Shan$^{1, 2, 3}$\thanks{E-mail: \url{hyshan@shao.ac.cn}},
Yiping Shu$^{4}$,
Nan Li$^{5}$,
Ji Yao$^{1}$,
Ran Li$^{5, 6, 2}$,
Xiaoyue Cao$^{2, 5}$,
\newauthor{Zizhao He$^{4}$,
Yin Li$^{7}$,
Eric Jullo$^{8}$,
Jean-Paul Kneib$^{9}$,
and Guoliang Li$^{4}$}
\\
$^{1}$Shanghai Astronomical Observatory, Chinese Academy of Sciences, Shanghai 200030, China\\
$^{2}$School of Astronomy and Space Science, University of Chinese Academy of Sciences, Beijing 100049, China\\
$^{3}$Key Laboratory of Radio Astronomy and Technology, Chinese Academy of Sciences, A20 Datun Road, Chaoyang District, Beijing, 100101, P. R. China\\
$^{4}$Purple Mountain Observatory, Chinese Academy of Sciences, Nanjing 210023, China\\
$^{5}$National Astronomical Observatory, Chinese Academy of Sciences, Beijing 100012, China\\
$^{6}$Institute for Frontiers in Astronomy and Astrophysics, Beijing Normal University, Beijing 102206, China\\
$^{7}$Department of Mathematics and Theory, Peng Cheng Laboratory, Shenzhen, Guangdong 518066, China\\
$^{8}$Aix Marseille Univ., CNRS, CNES, LAM, Marseille, France\\
$^{9}$Laboratory of Astrophysics, École Polytechnique Fédérale de Lausanne (EPFL), Observatoire de Sauverny, CH-1290 Versoix, Switzerland\\
}

\date{Accepted XXX. Received YYY; in original form ZZZ}

\pubyear{2025}

\begin{document}
\label{firstpage}
\pagerange{\pageref{firstpage}--\pageref{lastpage}}
\bibliographystyle{mnras}
\maketitle

\begin{abstract}

Strongly lensed supernovae (glSNe) provide a powerful, independent method to measure the Hubble constant, $H_{0}$, through time delays between their multiple images. The accuracy of this measurement depends critically on both the precision of time delay estimation and the robustness of lens modeling. In many current cluster-scale modeling algorithms, \textcolor{black}{all multiple images used for modeling} are simplified as point sources to reduce computational costs. In the first paper of the CURLING program, we demonstrated that such a point-like approximation can introduce significant uncertainties and biases in both magnification reconstruction and cosmological inference. In this study, we explore how such simplifications affect $H_0$ measurements from glSNe. We simulate a lensed supernova at $z=1.95$, lensed by a galaxy cluster at $z=0.336$, assuming time delays are measured from LSST-like light curves. The lens model is constructed using JWST-like imaging data, utilizing both \textsc{Lenstool} and a pixelated method developed in CURLING. Under a fiducial cosmology with $H_0=70\rm \ km \ s^{-1}\ Mpc^{-1}$, the \textsc{Lenstool} model yields \textcolor{black}{$H_0=69.91^{+6.27}_{-5.50}\rm \ km\  s^{-1}\ Mpc^{-1}$}, whereas the pixelated framework improves the precision by over an order of magnitude, \textcolor{black}{$H_0=70.39^{+0.82}_{-0.60}\rm \ km \ s^{-1}\ Mpc^{-1}$}. Our results indicate that in the next-generation observations (e.g., JWST), uncertainties from lens modeling dominate the error budget for $H_0$ inference, emphasizing the importance of incorporating the extended surface brightness of multiple images to fully leverage the potential of glSNe for cosmology.

\end{abstract}




\section{Introduction}

The Hubble constant ($H_{0}$) is a critical cosmological parameter that describes the expansion rate of the Universe. The value of $H_{0}$, however, is still under debate due to the significant discrepancy between early- and late-Universe measurements. Early-time observations of the cosmic microwave background \citep[CMB, ][]{planck20} yield a value of $H_{0}=67.4 \pm 0.5$ km/s/Mpc under the standard $\Lambda$ Cold Dark Matter ($\Lambda$CDM) model. In contrast, type \uppercase\expandafter{\romannumeral1}a supernovae (SNe \uppercase\expandafter{\romannumeral1}a) calibrated with distance ladders in the local Universe give $H_{0}=73.0 \pm 1.0$ km/s/Mpc \citep{riess22, zhang17, wong20, riess24}, remaining the problem unresolved as the `$H_{0}$ tension' \citep{abdalla22, peri22}.

Additional cosmological probes are necessary to resolve this 5$\sigma$ tension and determine the exact value of $H_{0}$. One promising method involves exploiting the gravitational lensing effect of \textcolor{black}{time-varying} sources. In strong lensing, a background source's light is deflected by the gravitational potential of an intervening mass, producing multiple images with distinct light paths. When the source is variable, these differences lead to measurable time delays between the images. The time delays are related to the so-called `time-delay distance', $D_{\Delta t}\equiv (1+z_{l}) D_{L}D_{S} / D_{LS}$, which is inversely proportional to $H_{0}$, \textcolor{black}{where $z_{l}$ is the redshift of the lens, $D_{L}$, $D_{S}$, and $D_{LS}$ are the angular diameter distances from the observer to the lens, from the observer to the source, and from the lens to the source, respectively.}

Time delay cosmology has been extensively used in lensed active galactic nuclei \cite[AGN, e.g., ][]{kochanek06, suyu17, courbin18, millon20b, ding21}, albeit the stochastic nature of AGN light curves and their longer variability time-scales pose challenges for precise measurement of $H_{0}$. As an alternative, lensed supernovae (glSNe) offer characteristic light curves and shorter timescales, making them more suitable for accurate time-delay measurements. However, glSNe are much rarer. The first discovery was SN Refsdal in the galaxy cluster MACS J1149.6+2223 \citep{kelly15}, whereas the idea of using the SN time delays to constrain $H_{0}$ dates back to 1964 \citep{refsdal64}. Among the seven confirmed glSNe \citep{kelly15, rodney21, chen22, frye24, pierel24b, goobar17, goobar23, goobar25} to date, only three have been utilized for $H_{0}$ measurement. \cite{grillo18} obtained $H_{0} = 69.8^{+5.3}_{-4.1}\rm \ km \ s^{-1}\ Mpc^{-1}$ from `SN Refsdal', while \cite{pascale24} derived $H_{0} = 75.4^{+8.1}_{-5.5}\rm \ km \ s^{-1}\ Mpc^{-1}$ using the `SN H0pe'. A more recent study of `SN Encore' yielded $H_{0} = 66.9^{+11.2}_{-8.1}\rm \ km \ s^{-1}\ Mpc^{-1}$ \citep{suyu25}, leaving significant uncertainties in $H_{0}$ inferences with glSNe.

The low rate of glSNe has prompted numerous efforts on the searching strategies \citep{shu18, craig21, huber21, magee23, nikki24}, while another bottleneck for glSNe time delay cosmography is the robustness of lens modeling \citep{grillo20, birrer22}, which is recognized as the dominant source of uncertainty ($\sim$ 5.5\%) over time delay measurement ($\sim 1.5\%$) in $H_{0}$ constraints \citep{kelly23}. Compared to galaxy-scale lenses, galaxy clusters have definite advantages, such as resolving the intractable mass-sheet degeneracy \citep{bradac04} when more than one family of multiple images are used for mass modeling, and longer time scales between image appearances that are easier to observe. Nevertheless, the complex mass distribution of clusters--featuring multiple dark matter clumps and hundreds of member galaxies--makes lens modeling more difficult, leading to higher uncertainties in cluster-based $H_0$ estimates \citep{suyu17, wong20, millon20a}. 

Frequently investigated systematic errors in cluster-scale lens modeling include the unaccounted line-of-sight (LoS) structures \citep{daloisio10, chirivi18}, the use of unified scaling relations across all member galaxies \citep{bergamini21}, the inaccuracies in the measurement of multiple image redshifts \citep{acebron17}. Aside from these, using only the positions of multiple images, instead of their extended surface brightness, can introduce systematic errors in lens modeling \citep{acebron24, cagan24}. This issue has been emphasized in the first paper of our CURLING \citep[ClUsteR strong Lens modelIng for the Next-Generation observations,][X24]{xie24} program, where we proposed a `pixelized model' framework. In this approach, lensing constraints are derived from the full surface brightness distribution of the lensed images, rather than just their positions.

\textcolor{black}{We expect to discover and study glSNe through repeated observations.} An ideal survey is the upcoming Large Synoptic Survey Telescope \citep[LSST, ][]{lsst}, which is poised to revolutionize time-domain astronomy with its wide field of view and deep imaging capabilities, covering the targeted regions every few nights. Another promising facility is the Multi-Channel Imager (MCI) aboard the China Survey Space Telescope \citep[CSST, ][]{csst18, zhan21}, which will offer unique advantages for glSNe detection. As a space-based instrument, MCI is free from atmospheric seeing, enabling better imaging compared to ground-based surveys. This is particularly valuable for observing the supernovae. MCI features wide wavelength coverage from NUV to NIR, extreme depth up to 30 magnitude, and a total exposure time exceeding 480 000 s (300 s $\times$ 1600 exposures). These capabilities make MCI especially well-suited for monitoring glSNe with high precision. In this study, we incorporate the observational characteristics of LSST and CSST-MCI into our simulations to assess their performance in enabling precise time-delay measurements for glSNe-based $H_0$ inference.

In this paper, we investigate the improvement of glSNe time delay cosmography with the pixelated modeling method. The work is organized as follows. In Section~\ref{section2}, we briefly introduce the theoretical framework of time-delay cosmography. We describe the simulated data used in this study, including the lens systems and the lensed supernova, in Section~\ref{section3}. In Section~\ref{section4}, we detail the measurement of time delays and the construction of lens models using both \textsc{lenstool} and our pixelized approach. Section~\ref{section5} presents the inferred values of $H_0$, comparing the results obtained from the two modeling methods. Finally, we summarize the work and discuss the findings in Section~\ref{section6}. Throughout the paper, we assume a flat $w$CDM cosmology ($\Omega_{\Lambda,0} = 0.7$, $\Omega_{\rm m}=0.3$, $h$ = 0.7) with a constant dark energy equation of state parameter $w = w_{0} = -1$.


\begin{table*}
\centering
\caption{Summary of the mock lens system based on MACS J0138.0-2155 at redshift $z=0.336$. The upper part describes the lensing mass components, which adopts the best-fit mass model of this cluster presented in \citet{rodney21} as input. The mock cluster includes a cluster-scale halo, a brightest cluster galaxy (BCG), three individual perturbers, and 24 additional cluster member galaxies. Parameters for the member galaxies follow unified scaling relations. For each component, the table lists the position to the cluster center ($\Delta \alpha$, $\Delta \delta$), ellipticity $e$, position angle $\theta$, core radius $\rm r_{core}$, cut radius $\rm r_{cut}$, and velocity dispersion $\sigma$. The lower part presents the lensed sources, each modeled using a S$\rm\acute{e}$rsic profile characterized by the position, semi-major and semi-minor axes ($a$, $b$), position angle ($\theta$), appararent magnitude, and source redshift $z_{s}$. Source 2 corresponds to the strongly lensed supernova.}
\setlength{\tabcolsep}{10pt} 
\small 
    \begin{tabular}{c c c c c c c c}
 \hline
 \multicolumn{8}{l}{MACS0138-like,  $z = 0.336$}\\
 \hline
          Component & $\Delta \alpha$  & $\Delta \delta$  & \it{e} & $\theta$ & $\textcolor{black}{{\rm r}}_{\rm core}$ & $\textcolor{black}{{\rm r}}_{\rm cut}$ & $\sigma$ \\
           & [$^{\prime \prime}$] & [$^{\prime \prime}$] & & [deg] & [kpc] & [kpc] & [km/s] \\
 \hline

         Cluster Halo  & -0.7 & -1.2 & 0.81 & 114.9 & 31.0 & 1000.0 & 446.0 \\
         BCG & 0.1 & -0.1 & 0.52 & -41.1 & 0.15 & 136.0 & 700.0 \\
         Perturber 1 & 20.04 & -15.11 & 0.49 & 86.2 & 0.15 & 25.5 & 152.0 \\
         Perturber 2 & -5.0 & 6.9 & 0.06 & 4.4 & 0.15 & 12.0 & 23.0 \\
         Perturber 3 & 0.12 & -18.25 & 0.24 & -63.1 & 0.15 & 6.0 & 110.0\\
         scaling relations & N(gal) = 24 & ${\rm m}^{\rm ref}$ = 19.5 & $\it r_{\rm core}^{\rm ref}$ = 0.15 kpc & $\it r_{\rm cut}^{\rm ref}$ = 45.0 kpc& $\sigma^{\rm ref}$ = 158.0 km/s& & \\
    \hline
 \multicolumn{8}{l}{Sources}\\
    \hline
        Component & $\Delta \alpha$  & $\Delta \delta$  & \it{a} & \it{b} & $\theta$ & mag & $z_{\rm s}$ \\
        & [$^{\prime \prime}$] & [$^{\prime \prime}$] & [kpc] & [kpc] & [deg] & &  \\
    \hline
        Source 1 & 3.05 & -1.47 & 2.17& 3.47 & 0.0 & 15.5 & 1.95 \\
        Source 2 (SN) & 3.1 & -1.73 & / & / & 0.0 & 20.8 & 1.95\\
        Source 3 & -2.0 & 4.01 & 1.88 & 1.13 & 0.0 & 18.5 & 0.7336 \\
 \hline
    \end{tabular}
\label{table1}
\end{table*}

\section{Time delay cosmography} \label{section2}

In this section, we briefly introduce the theory of time delay cosmology. For a comprehensive description, we refer the reader to \cite{blandford92, schneider92, treu16, meneghetti22} and references therein.

When light from a distant source is deflected by a gravitational lens, it travels along different paths and through varying gravitational potentials, resulting in a delay in arrival time compared to the unlensed case. The total time delay consists of two components: 
\begin{equation} \label{eq1}
	t = t_{\rm grav} + t_{\rm geom}.
\end{equation}
The first term is the gravitational time delay,
\begin{equation} \label{eq2}
    t_{\rm grav} = \int \frac{dz}{c'} - \int \frac{dz}{c} = -\frac{D_{L}D_{S}}{D_{LS}}\frac{1}{c}\hat{\Psi},
\end{equation}
caused by the intervening gravitational potential $\hat{\Psi}$. The second term denotes the geometrical time delay,
\begin{equation} \label{eq3}
    t_{\rm geom} = \frac{\Delta l}{c} = \frac{1}{2c} (\vec{\theta}-\vec{\beta})^2 \frac{D_{L}D_{S}}{D_{LS}},
\end{equation}
where $\Delta l$ represents the extra path of the light in the presence of the lens. By \textcolor{black}{substituting} Equations \ref{eq2} and \ref{eq3} into Equation \ref{eq1}, and accounting for the time dilation by the factor (1+$z_{\rm L}$), we obtain the time delay at image position $\vec{\theta}$:
\begin{equation} \label{eq4}
    t(\vec{\theta}) = \frac{1+z_{\rm L}}{c}\frac{D_{L}D_{S}}{D_{LS}} [\frac{1}{2} (\vec{\theta} - \vec{\beta})^2 - \hat{\Psi}(\vec{\theta})] = \frac{D_{\Delta t}}{c} \tau(\vec{\theta}).
\end{equation}
Here, the term $D_{\Delta t} = (1+z_{\rm L}) \frac{D_{L}D_{S}}{D_{LS}}$ is defined as the time-delay distance, which encapsulates three angular diameter distances, $D_{L}$ from the observer to the lens, $D_{S}$ from the observer to the source, and $D_{LS}$ from the lens to the source. According to the definition of the angular diameter distance,
\begin{equation}
    D(z_{1},z_{2}) = \frac{c{H}_{0}^{-1}}{1+z_{2}}\int_{z_{1}}^{z_{2}}dz [\Omega_{\rm m}(1+z)^{3}+\Omega_{\rm X}(1+z)^{3(w_{\rm X}+1)}]^{-1/2},
\end{equation}
the time-delay distance $D_{\Delta t}$ is inversely proportional to $H_{0}$ and weakly dependent on other cosmological parameters. The term $\tau(\vec{\theta})$ in Equation~\ref{eq4} is the Fermat potential, which can be determined through lens modeling.

Though the arrival time of one light ray is not detectable, the time difference between two lensed images $i$ and $j$ is measurable and given by:
\begin{equation} \label{eq6}
    \Delta t_{ij} = \frac{D_{\Delta t}}{c} [\tau(\vec{\theta_{i}}) - \tau(\vec{\theta_{j}})].
\end{equation}
Therefore, if the time delay $\Delta t_{ij}$ is measured from observations and the Fermat potential is accurately inferred from lens modeling, the value of $H_{0}$ can be well constrained.


\section{Simulations} \label{section3}

The primary aim of this study is to compare the use of point-like versus extended multiple images as lensing constraints, in order to assess how extended surface brightness information can improve the inference of the Hubble constant $H_0$ from strongly lensed supernovae in galaxy clusters. To focus on this comparison, we use mock data instead of real observations, which allows us to avoid additional challenges, such as light contamination on the extraction of arcs, deviations between analytical models and real mass distributions, that would complicate the modeling of extended sources. 

In this section, we describe the construction of the simulated data used in this work. Simulating a lensed supernova involves two main components: defining the mass distribution of the lensing galaxy cluster and generating multiple images of the lensed sources which will serve as `observational' constraints for lens modeling; specifying the properties of the background supernova.

\begin{figure*}
	\includegraphics[width=0.98\textwidth]{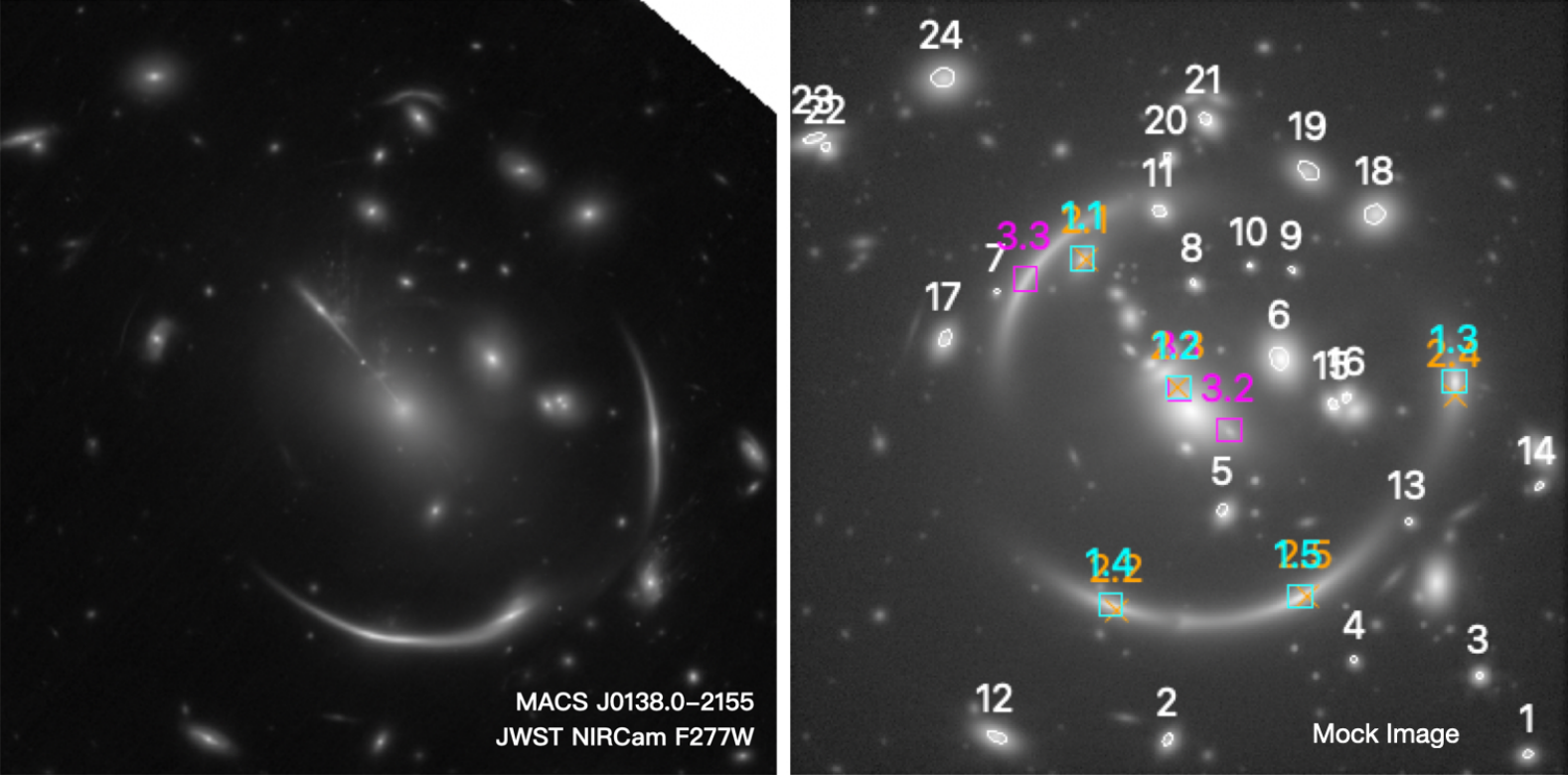}
    \caption{Left: JWST NIRCam (F277W) Observation on the lensing cluster MACS J0138.0-2155. Right: Mock image of the simulated cluster used in this work. Member galaxies, multiple images as lensing constraints, and lensed supernova images are marked with white ellipses, cyan/magenta boxes, and orange crosses, respectively. The field-of-view (FoV) is $60^{\prime\prime}\times 60^{\prime\prime}$.}
    \label{fig1}
\end{figure*}

\subsection{The strong lensing system: MACS J0138.0-2155}\label{subsection31}

To create a realistic mock lensing cluster, we simulate the field of MACS J0138.0-2155, a galaxy cluster of particular interest due to the discoveries of two lensed supernovae, SN Requiem and SN Encore, located in the same host galaxy \citep{rodney21, pierel24b}. Making the mock image for this cluster field involves several key steps.

\begin{enumerate}
    \item We generate the galaxies in the field, which includes not only cluster member galaxies but also foreground and background galaxies along the line of sight. To ensure a reliable spatial and photometric distribution, we perform source detection and photometry using \textsc{Photutils} \citep{photutils} on the JWST NIRCam F277W image (PI: Justin Pierel, Proposal ID: 6549) of MACS J0138.0-2155. From this, we extract the positions, shapes, and fluxes of all detected galaxies. Each galaxy is then simulated using a S$\rm \acute{e}$rsic profile, adopting the measured morphological parameters and flux.
    \item We define the mass distribution of the cluster lens system by adopting the best-fit lens model from \cite{rodney21} as the input cluster mass, as detailed in Table~\ref{table1}. The model includes a cluster-scale dark matter halo, a brightest cluster galaxy (BCG), three perturbers close to the multiple images, \textcolor{black}{and member galaxies. While \cite{rodney21} modeled 32 members, our simulation includes only 24, as we restrict the field-of-view to $60^{\prime\prime} \times 60^{\prime\prime}$. This clipping reduces computational cost while preserving the strong-lensing region at the core of the cluster, ensuring that the critical information for lens modeling are retained.} Each component is described using a pseudo-isothermal elliptical mass distribution \citep[PIEMD, ][]{eliasdottir07}, 
    \begin{equation}
        \rho(r) = \frac{\rho_{0}}{(1+\frac{r^2}{\textcolor{black}{{\rm r}}_{\rm core}^2})(1+\frac{r^2}{\textcolor{black}{{\rm r}}_{\rm cut}^2})},
    \end{equation}
    where $\textcolor{black}{{\rm r}}_{\rm core}$, $\textcolor{black}{{\rm r}}_{\rm cut}$ are the core radius and cut radius, respectively. A PIEMD profile is thereby described by the parameter set \{ $x$, $y$, $e$, $\theta$, $\rm r_{core}$, $\rm r_{cut}$, $\sigma$ \}, where $x$, $y$ describe the center of the potential, $e$ and $\theta$ are the ellipticity and position angle, respectively, and $\sigma$ represents the velocity dispersion of the potential, which is related with $\rho_{0}$ via $\sigma^2 = \frac{4}{3}G\pi \rho_{0} \rm r_{core}^2 r_{cut}^3/[\rm (r_{cut} - r_{core})(r_{cut} + r_{core})^2]$. Core radius, cut radius, and velocity dispersion of each member galaxy follow scaling relations based on its magnitude \citep{granata22}:
    \begin{align}\label{scaling}
    \begin{split}
        \left \{
        \begin{array}{ll}
        \rm  \sigma^{gal}_{\it{i}}=\sigma^{ref} 10^{0.4 \frac{m^{ref}-m_{\it{i}}}{\alpha}}\\
        \textcolor{black}{{\rm r}}^{\rm gal}_{\rm core,\it{i}}=\textcolor{black}{{\rm r}}_{\rm core}^{\rm ref}10^{0.4 \frac{m^{\rm ref}-m_{\it{i}}}{2}}\\
        \textcolor{black}{{\rm r}}^{\rm gal}_{\rm cut,\it{i}}=\textcolor{black}{{\rm r}}_{\rm cut}^{\rm ref}10^{0.4 \frac{2(m^{\rm ref}-m_{\it{i}})}{\beta}},
        \end{array}
        \right.
    \end{split}
    \end{align}
    where the typical values $\sigma^{\rm ref}$, $\textcolor{black}{{\rm r}}_{\rm core}^{\rm ref}$, $\textcolor{black}{{\rm r}}_{\rm cut}^{\rm ref}$, and $m^{\rm ref}$ are 158 $\rm km\ s^{-1}$, 0.15 kpc, 45 kpc, and 19.5 mag.
    \item We generate the lensed arcs based on the mass distributions obtained in step (ii). To do so, we first estimate the properties of the sources, including their positions, shapes, and fluxes, assuming a S$\rm \acute{e}$rsic profile for each. Using these sources, we perform ray tracing through the cluster's \textcolor{black}{gravitational} lensing potential to create mock arcs, and adjust the source properties to find the best arcs that resemble those observed in the real cluster. The properties of the two sources corresponding to the arcs and the supernova are  summarized in Table~\ref{table1}.
    \item We model the intra-cluster light \citep[ICL, ][]{zwicky37, demaio18, chenxk22, contini23}, which accounts for the diffuse light of individual stars that are not associated with any galaxy. The ICL is represented by a smooth component following a Navarro-Frenk-White \citep[NFW, ][]{nfw} profile, of which its mass is determined according to the total cluster stellar mass $M_{*}=9.26\times10^{12} \rm M_{\odot}$, and a concentration parameter $c=3.88$, based on the concentration-mass relation of \cite{child18}. More details for the generation of mock ICL distribution can be found in \cite{chenxk22}.
    \item Finally, we simulate observational effects to create the mock image. The mock image is convolved with a JWST NIRCam Point Spread Function (PSF) generated using \textsc{webbpsf} \citep{webbpsf}, and we add noise to reflect observing conditions. This includes both Poisson noise from the sources and sky background, as well as Gaussian noise from detector readout. The total noise level in each pixel is computed \citep{acs2024, cao18} using:
    \begin{equation} \label{eq8}
        \sigma = \frac{\sqrt{N_{\rm pix}\times (S_{\rm img} + \ S_{\rm sky})\times t_{\rm expo} + N_{\rm expo}\times R^2}}{t_{\rm expo}},
    \end{equation}
    where $S_{\rm sky}$ is the sky brightness in unit of $e^{-}\ \rm s^{-1}\ pixel^{-1}$, converted from $m_{\rm sky}=29.5 \rm \ mag\ arcsec^{-2}$, given a zero magnitude of 27.295 in JWST F277W filter, $N_{\rm pix}$ is the number of pixels of the object, the readout noise is $R=15.88\ e^{-}$, the total exposure time is $t_{\rm expo}=7\ 537$ s, and the number of exposures is $N_{\rm expo}=9$.
\end{enumerate}
The mock image, generated through the process detailed above, is presented in Figure~\ref{fig1} alongside the real JWST/NIRCam F277W observation for visual comparison.

\begin{figure}
	\includegraphics[width=0.49\textwidth]{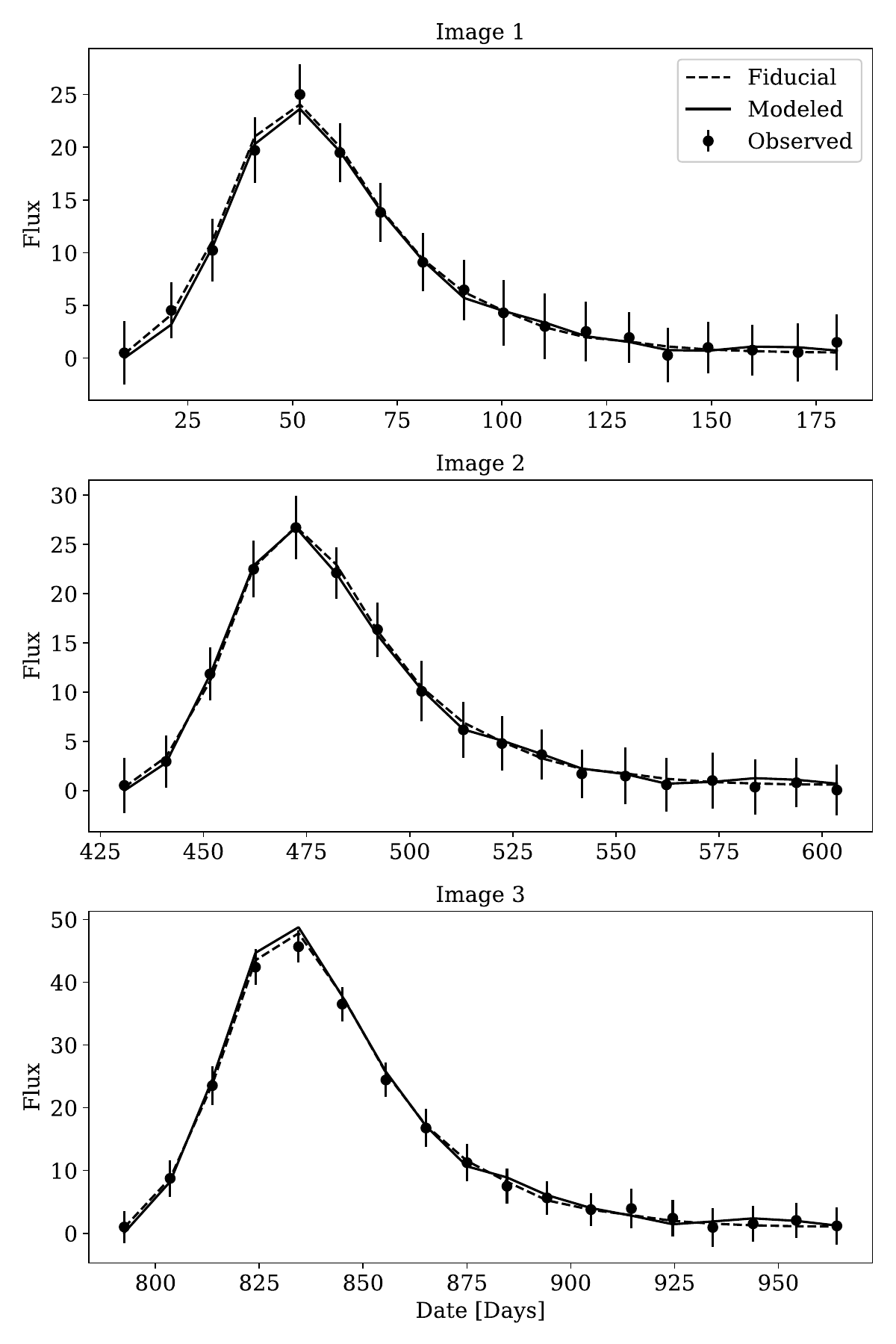}
    \caption{\textcolor{black}{Light} curves of the three multiple images of the simulated supernova under the LSST observational configuration. Each panel corresponds to one of the SN images. The dashed lines represent the theoretical light curves computed from the intrinsic supernova model and the associated lensing magnification and arrival time. Data points with error bars show the measured fluxes, incorporating observational effects from LSST. Solid lines indicate the best-fit light curves obtained using the \textsc{sntd} package.}
    \label{fig2}
\end{figure}

\subsection{The supernova: Requiem}\label{subsection32}

Two aspects need to be taken into account for simulating the lensed supernova: the intrinsic properties of the supernova and the effects of gravitational lensing.

For the supernova itself, we model a Type \uppercase\expandafter{\romannumeral 1}a event inspired by SN Requiem in MACS J0138.0-2155 \citep{rodney21}, located at a source redshift of $z_{s}=1.95$. The SN light curve is generated using the \texttt{salt2} model \citep[version \texttt{T23}, ][]{taylor23} provided by \textsc{SNCosmo}, adopting typical parameters from the literature \citep{Kenworthy2021}, $x_1=0.4$ for the light curve shape, $c=-0.05$ for the color, and $t_0=0$ days as the observer-frame time corresponding to the source's phase = 0. The SN is simulated with a peak absolute magnitude of -19.2 AB mag in B band, which is a typical value for Type \uppercase\expandafter{\romannumeral 1}a supernova \citep{phillips93, riess98}.

For lensing, the SN is placed with an offset of $\Delta x=0.43\rm \ kpc$ and $\Delta y=2.24\rm \ kpc$ from the host galaxy's center. Computations through the cluster potential described in Section~\ref{subsection31} \textcolor{black}{yield} five lensed images of the SN (orange crosses in Figure~\ref{fig1}) \textcolor{black}{with respective magnifications and arrival times of (1.33, 3.58, $3.88\times10^{-4}$, 3.23, 6.45) and (5 968.85, 421.97, 9 318.08, 0, 781.98) days after the first image appears}. Given the low magnifications and late arrival times of \textcolor{black}{SN image 2.1 and 2.3, we focus on the other three for time delay measurements, which we label as SN image 2.2, 2.4, and 2.5. These images exhibit time delays relative to the earliest image 2.4 of $\Delta t_{42}=421.97$ days and $\Delta t_{45}=781.98$ days}.

Based on the original SN light curve, we compute the theoretical light curves for these three lensed images and show them as the dashed lines in Figure~\ref{fig2}. To simulate observational conditions, we assume that the supernova images are monitored by LSST with a Gaussian-distributed cadence (a mean of 10 days and a standard deviation of 0.5 day), a seeing of 0.68$^{\prime\prime}$ \citep{nikki24}, and noise as described in Equation~\ref{eq8} with sky magnitude $m_{\rm sky}=22.61 \rm \ mag\ arcsec^{-2}$, zero point magnitude in g-band $m_{\rm zp}=26.68$, $t_{\rm expo}=30\ s$, $N_{\rm expo}=2$, and $R=10\ e^{-}$. 

\textcolor{black}{After generating the simulations, we replicate observational procedures to extract the light curves. At each epoch, we subtract a reference frame without SN flux from a science frame containing the corresponding flux, with both sets of images produced under identical observing conditions. Force photometry is then performed at the known SN image positions on the difference images, measuring fluxes within an aperture matched to the PSF size. These measurements yield the data points shown in Figure~\ref{fig2}, which form the basis for deriving the time delays of the lensed SN images, as depicted in Section~\ref{subsection42}.}



\begin{figure*}
    \centering
    \begin{subfigure}[b]{0.49\textwidth} 
        \centering
        \includegraphics[width=\linewidth]{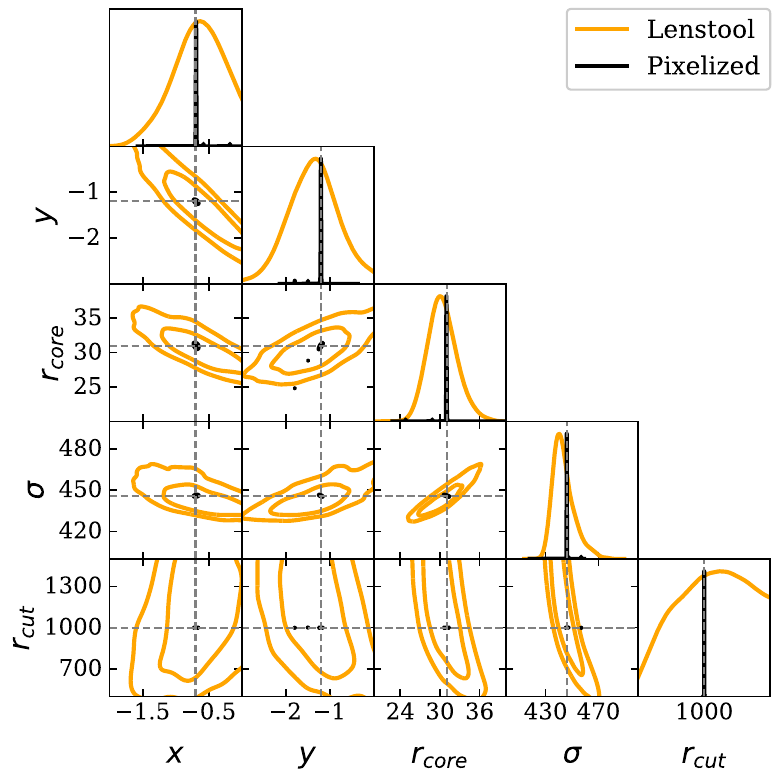} 
    \end{subfigure}
    \hfill 
    \begin{subfigure}[b]{0.49\textwidth}
        \centering
        \includegraphics[width=\linewidth]{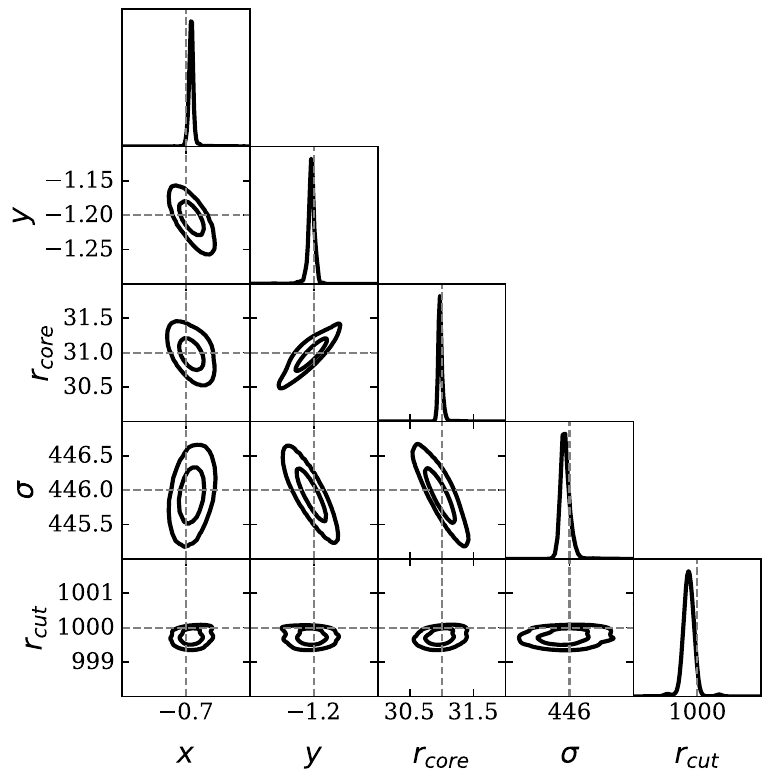} 
    \end{subfigure}
    \caption{Posterior distributions \textcolor{black}{(1-$\sigma$ and 2-$\sigma$ confidence levels)} of the lens mass parameters for the main halo. The left panel compares the results obtained using \textsc{Lenstool} (orange) and the pixel-based modeling approach (black). The right panel provides a zoomed-in view of the pixel-based modeling posteriors.} 
    \label{fig3}
\end{figure*}

\section{Methods} \label{section4}

\subsection{Lens modeling}

For the purpose of assessing whether the pixel-based modeling approach can improve the $H_0$ inference, we model the simulated galaxy cluster introduced in the previous section by using two techniques: a traditional one that uses only the positions of multiple images as observational constraints, and a new one developed in our earlier work \citep{xie24}, which fits the extended surface brightness distribution of lensed arcs. Both methods are applied under a simulated JWST/NIRCam observation setup and make use of the same two families of multiple images as constraints.

Due to computational limitations, only the parameters of the primary cluster-scale halo are allowed to vary, while the remaining model parameters are fixed to their inputs. Systematics such as uncertainties in redshift measurements, the presense of line-of-sight structures are not included.

Cluster-scale lens modeling has historically treated multiple images of background sources as point-like constraints, fitting lens models by minimizing the discrepancies between the observed and predicted positions of lensed images, while ignoring morphological and photometric information. Here, we use \textsc{Lenstool} \citep{kneib96, jullo07} as a representative of this method. The optimization is performed by minimizing a $\chi^2$ defined on the image plane:
\begin{equation}\label{eq9}
    \chi_{i} ^{2} =\sum_{j=1}^{{\rm n}_{i}} \frac{[\vec{\theta} _{\mathrm obs}^{j}-\vec{\theta}^{j}({\it p})] ^{2} }{\sigma_{ij}^{2}},
\end{equation}
where $\vec{\theta}_{\mathrm obs}^{j}$ and $\vec{\theta}^{j}({\it p})$ are the observed and modeled positions of image $j$, respectively. The positional uncertainty $\sigma_{ij}$ is uniformly assumed to be $0.1^{\prime\prime}$, consistent with the value used for JWST strong lensing cluster modeling \citep[e.g., ][]{bergamini23}. The posterior distribution of the lens parameters \textit{p} is sampled using the \textsc{BayeSys} MCMC sampler \citep{skilling05} implemented within \textsc{Lenstool}.

Pixel-based lens modeling becomes increasingly significant with the availability of high-resolution imaging \citep{acebron24, urcelay25, cao25, ding25}, since the traditional lens modeling approach discards the rich information contained in the surrounding pixels. As a result, much of the morphological detail and flux distribution within the lensed arcs is ignored, leading to a substantial loss of constraining power. Furthermore, this method relies on the assumption that positional uncertainties follow a Gaussian distribution, while the complex and arc-like morphology of these sources violates this assumption, potentially introducing biases into the inferred lens model parameters. 

To overcome the limitations in traditional lens modeling techniques, we implement a more sophisticated modeling framework developed in X24, which fits the observed surface brightness distributions of the lensed images on a pixel-by-pixel basis, enabling leveraging full information of the arcs and more precise parameter inference. The modeling procedure consists of the following steps. First is the arc pixel selection, we identify high-signal-to-noise regions of the arcs by selecting pixels centered around the flux peaks of each lensed image. Only pixels with fluxes greater than 60 times the mean value of the error map are included, ensuring reliable photometric constraints. $23,000$ pixels are selected as the arcs in this simulated cluster. The next step is forward modeling, for each proposed set of model parameters \textit{p}, we generate the corresponding image distributions via lens equation. Bayesian inference is finally applied, where the $\chi^2$ over all selected image pixels is defined as:
\begin{equation}
    \chi ^{2} =\sum_{i=1}^{{\rm n_{pix}}} \frac{[{S} _{\mathrm obs}^{i}-{S}^{i}({\it p})] ^{2} }{\sigma_{i}^{2}},
\end{equation}
where $S_{\mathrm obs}^{i}$ is the observed surface brightness at pixel $i$, $S^{i}({\it{p}})$ is the model-predicted surface brightness given model parameters ${\it p}$, and $\sigma_{i}$ is the associated error. The noise model is based on a JWST/NIRCam configuration as mentioned in Section~\ref{subsection31}.

Figure~\ref{fig3} presents the posterior distributions of the primary halo parameters obtained from the two modeling approaches. The left panel compares the results from \textsc{Lenstool} (shown in orange) with those from the pixel-based modeling (shown in black). A zoomed-in version of the pixel-based modeling posteriors is presented in the right panel. The true parameter values used in the simulation are indicated by dashed lines. As previously emphasized in X24, the pixel-based modeling yields significantly improved constraints, the posterior contours are narrower, demonstrating its precision in parameter estimation.

\subsection{Time delay measurement}\label{subsection42}

We employ the open-source Python package \textsc{sntd} \citep[SuperNova Time Delays, ][]{sntd, pierel19} to measure the time delays between the multiple images of the simulated lensed supernova (see Section~\ref{subsection32}). Built on top of the well-established \textsc{sncosmo} framework \citep{barbary16}, \textsc{sntd} enables time delay inference by fitting the light curves of the multiple images. We adopt the `parallel' fitting approach, which allows for fitting the light curve of each image in parallel, provided that the light curves are well sampled before and after the brightness peaks. Given the fact that the various light curves for the multiple images are generated from the same SN explosion, the parameters for fitting the light curves are divided into two categories, $\theta_{\rm SN}$, including parameters that describe the SN properties (e.g., redshift, type, light curve shape, etc), and $\theta_{\rm L}$, containing parameters affected by lensing (i.e., the magnification effect and changes in arrival times). The parameter space is explored using nested sampling, \textsc{sntd} first constrains $\theta_{\rm SN}$ using all light curves jointly, and then derives $\theta_{\rm L}$ for each image individually.

We fit the simulated light curves using the \texttt{salt2} model, consistent with that used in the simulation in Section~\ref{subsection32}. The free parameters include the light curve shape parameters $x_0$ and $x_1$, the color parameter $c$, and the time of maximum brightness $t_0$ in the observer frame. The supernova redshift is assumed to be known, and microlensing effects are not included in the simulations or fitting.

The solid lines in Figure~\ref{fig2} present the results of the light curve fitting, for the three images 1, 2, and 3 from top to bottom. \textsc{sntd} fits the `LSST-observed' fluxes (indicated as error bars) with high accuracy, as illustrated by the solid lines in Figure~\ref{fig2}. Although these observed fluxes deviate from the theoretical light curves (dashed lines) due to the effects of seeing and observational noise, we find that \textsc{sntd} still delivers robust and precise time delay measurements: \textcolor{black}{$\Delta t_{42, \rm \ obs}=421.18^{+2.16}_{-2.21}$} days, \textcolor{black}{$\Delta t_{45, \rm \ obs}=781.51^{+1.53}_{-1.73}$} days, in excellent agreement with the theoretical time delays of $\Delta t_{42}=421.97$ days and $\Delta t_{45}=781.98$ days. The deviations in measured fluxes due to seeing do not significantly affect the inferred delays, highlighting the robustness of light curve shape information in time delay measurements.


\begin{figure}
	\includegraphics[width=0.51\textwidth]{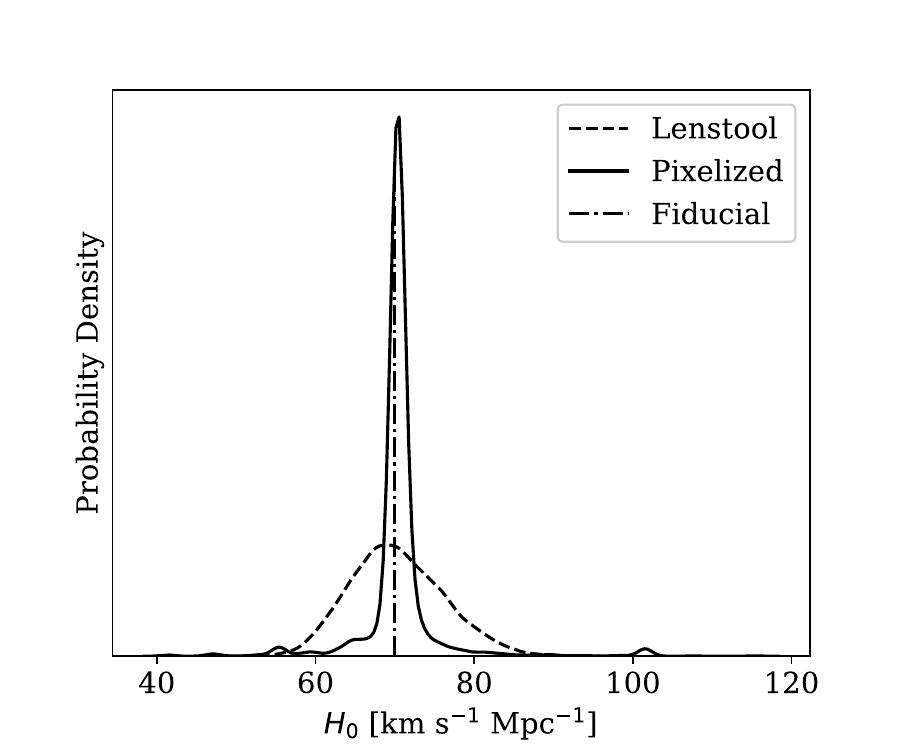}
    \caption{\textcolor{black}{Inferred} $H_{0}$ from lens model obtained with \textsc{lenstool} (dashed) and pixelized method (solid curve). The fiducial value of $H_{0, \rm \ fid}$ is marked as the vertical dash-dotted line.}
    \label{fig4}
\end{figure}


\section{Results} \label{section5}

Based on the derived lens models and the time delays for the simulated lensed supernova in our analyses, we are able to constrain the value of the Hubble constant $H_0$. The lens models are constructed assuming a fiducial value of $H_0=70\ \rm km\ s^{-1}\ Mpc^{-1}$, yielding fiducial time delays $\Delta t_{i, j}^{\rm fid}$ for each image pair ($i, \ j$). Given that the time delay between lensed images is inversely proportional to $H_0$ (Equation~\ref{eq6}), we can rescale the fiducial model-predicted time delays to infer the actual value of $H_0$ by comparing them to the observed time delays $\Delta t_{i,j}^{\rm meas}$. The measured time delay at its corresponding $H_0$ is given by:
\begin{equation}
    \Delta t_{i, j}^{\rm meas}(H_0)=\Delta t_{i,j}^{\rm fid}\times \frac{70\ \rm km\ s^{-1}\ Mpc^{-1}}{H_0}.
\end{equation}
By comparing these model-predicted time delays to the measured values for all image pairs, constraint on the value of $H_0$ is obtained assuming that the lens model accurately captures the underlying mass distribution \citep{pascale24, napier23}.

All other cosmological parameters are fixed at their true values, as our goal is to isolate and examine the impact of lens modeling accuracy on the inferred value of $H_0$. The constraints on $H_0$ derived from the lens models using both methods are shown in Figure~\ref{fig4}. We obtain \textcolor{black}{$H_0=69.91^{+6.27}_{-5.50}\rm \ km\  s^{-1}\ Mpc^{-1}$} (\textsc{Lenstool}, dashed curve) and a significantly tighter constraint \textcolor{black}{$H_0=70.39^{+0.82}_{-0.60}\rm \ km \cdot s^{-1}\cdot Mpc^{-1}$} (pixel-based modeling, solid curve). Both methods produce robust results under ideal conditions, while pixel-based approach offers an order of magnitude enhancement in precision. 

This enhancement is particularly important. In current studies using cluster-lensed supernovae, uncertainties in $H_0$ are dominated by lens modeling errors ($\sim$ 5.5\%, whereas uncertainties from time-delay measurements are much smaller ($\sim$ 1.5\%) \citep{kelly23}. Our findings suggest that adopting pixel-based modeling can substantially improve cosmological precision by delivering more accurate and constrained lens models.

\begin{figure}
	\includegraphics[width=0.49\textwidth]{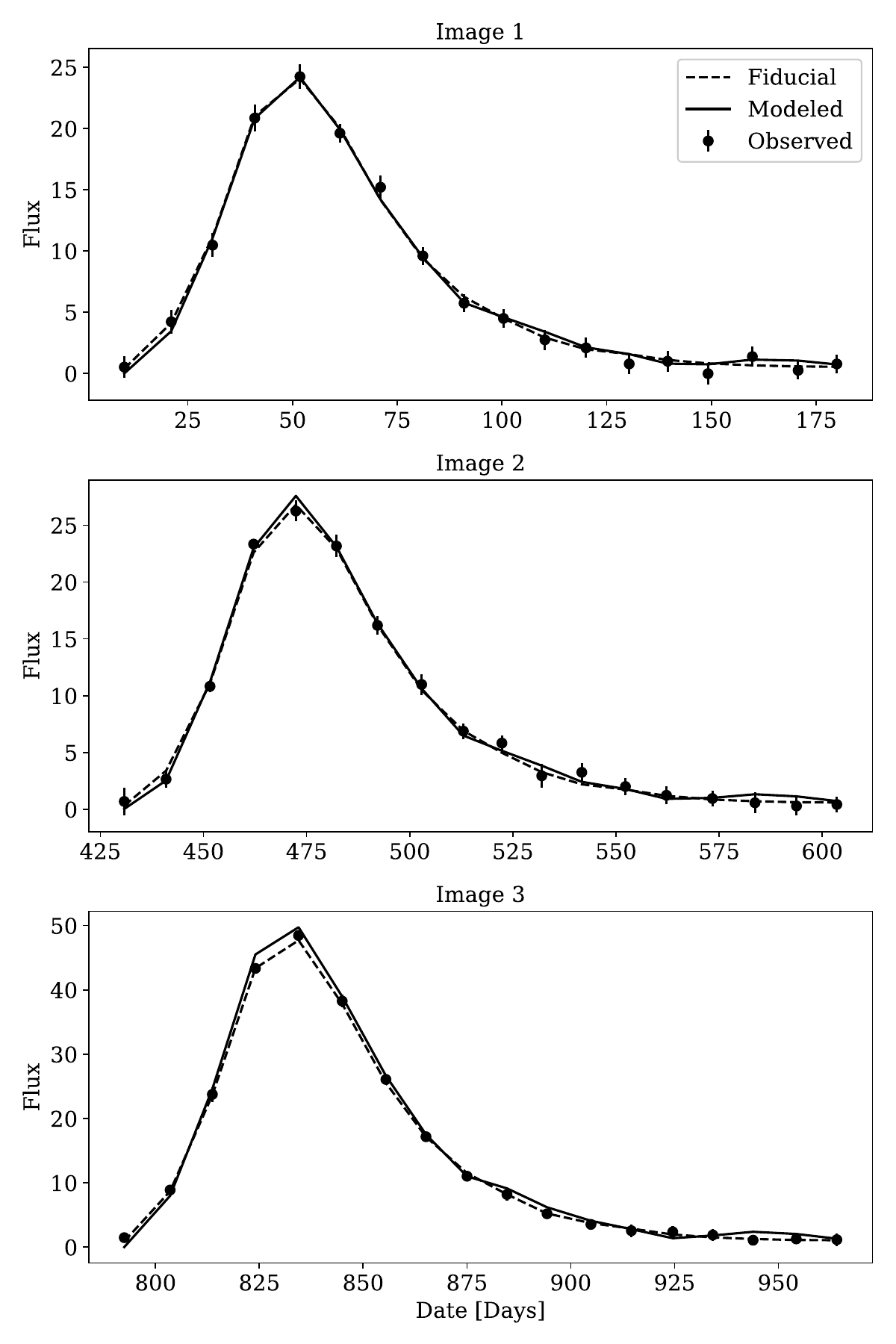}
    \caption{\textcolor{black}{Light} curves of the three multiple images of the simulated supernova under the CSST-MCI observational configuration. The line styles are consistent with those in \textcolor{black}{Figure~\ref{fig2}}.}
    \label{fig5}
\end{figure}

\section{Discussion and conclusion} \label{section6}

The glSNe are regarded as a promising one-step probe for measuring the distance scale of the Universe, potentially resolving  the 5$\sigma$ $H_{0}$ tension between the early- and late- time observations. Nevertheless, the robustness of lens modeling is a critial factor in achieving reliable constraints on $H_{0}$. As the second paper in the CURLING program, this work investigates the impact of incorporating pixel-level information from multiple images into lens modeling on the $H_{0}$ constraints, leveraging time delays between glSN multiple images.

We simulate a system analogous to the multiply imaged supernova `Requiem', discovered in the galaxy cluster MACS J0138.0-2155, and perform lens modeling using two approaches: the traditional parametric method implemented in \textsc{Lenstool}, which relies on the positions of multiple images, and a pixel-based method that incorporates the surface brightness distribution across image pixels. Our results show that the pixel-based approach yields a more robust and precise reconstruction of the lensing mass distribution. On the other hand, we measure the time delays between the SN images using the public package \textsc{sntd}, which fits the simulated light curves. Using the inferred time delays and lens models, we then derive $H_0$. While traditional modeling provides reasonable constraints, the pixel-based approach significantly improves precision, highlighting the value of using full surface brightness information in lens modeling.

To investigate the contribution of lens modeling uncertainties to $H_0$ inference, we compare the two modeling methods while fixing time delay uncertainties based on the LSST observing setup. Nonetheless, uncertainties in $H_0$ arise from both time delay measurements and lens modeling. To assess the impact of improved time delay precision, we consider an alternative observational configuration: the CSST-MCI, a space-based instrument which can avoid the influence from atmospheric seeing effects and offer high quality imaging data (observational parameters: sky magnitude $m_{\rm sky}$ = 21.68 mag $\rm arcsec^{-2}$, zero point magnitude in CBU band: $m_{\rm sky}$ = 26.373 mag, readout noise $R=10\ e^{-}$, total exposure time $t_{\rm expo}=300$ s, number of exposures $N_{\rm expo}=2$, and a pixel scale of 0.05$^{\prime\prime}$). The fitted light curves for the three supernova images observed by MCI are shown in Figure~\ref{fig5}. Using \textsc{sntd}, we obtain time delays of \textcolor{black}{$\Delta t_{42}=421.74^{+0.63}_{-0.64}$} days, and \textcolor{black}{$\Delta t_{45}=781.74^{+0.62}_{-0.70}$} days. These measurements yield \textcolor{black}{$H_0=69.88^{+6.25}_{-5.49}\ \rm km\ s^{-1}\ Mpc^{-1}$} with \textsc{Lenstool}, and \textcolor{black}{$H_0=69.84^{+0.11}_{-0.10
}\ \rm km\ s^{-1}\ Mpc^{-1}$} using the pixel-based modeling. Despite the significantly improved time delay precision with MCI compared to LSST, the constraint on $H_0$ with \textsc{Lenstool} remains similar, further confirming that lens modeling with this method dominates the error budget in $H_0$ inference. However, once lens modeling uncertainties are reduced, the high-precision time delay measurements from MCI, compared to those from LSST, can be fully leveraged to achieve much more stringent constraints on $H_0$.

Microlensing from passing stars or other small objects can perturb the time-delay measurements for glSNe, albeit it is largely mitigated comparing with strongly lensed AGN \citep{bonvin19, tie18} due to the compact sizes of SNe and their characteristic light curves. Microlensing introduces chromatic flux variations \citep{dobler06} that can distort the intrinsic SN brightness changes, making it challenging to accurately match corresponding features in the light curves of different multiple images. In order to avoid such variations, \cite{pierel24a} proposed to use colored light curves. Additionally, if the flux ratio of two images is taken into account in the time delay measurement, microlensing can introduce perturbation to the magnification, leading to a few tenths of magnitude bias, further complicating the time delay analysis. While we have not accounted for these effects in this work, we note the importance and will incorporate microlensing into our framework in future work to improve the accuracy and reliability.

We have made several simplifications for modeling the lensing cluster to ensure the time-consuming pixelized algorithm working: 1) we use the same form of analytical potentials for the simulation and fitting, incorporating no other systematic errors but instrumental and Poisson noises in the image; 2) we treat the parameters of the main halo as free parameters in the lens modeling, while assuming that the remaining potentials are well constrained already; 3) we ignore the complex structures of sources and masses on the line of sight, which can worsen the modeling results \citep{li21}; 4) we use only two families of multiple images as the observables, whereas more than tens of lensed sources could be identified thanks to the capabilities of deep imaging and spectroscopic follow-up observations. Since our aim here is to investigate the enhancement of our new framework with the pixelized images for lens modeling, a careful consideration for the realistic simulation of the lensing cluster is beyond the scope, while we believe that making the pixelized algorithm applicable to real observations is vital. With the aid of the promising GPU-based \textsc{JAX} framework \citep{jax2018github} combined with the high-performance samplers, e.g., \textsc{numpyro} \citep{numpyro}, our method is anticipated to be accelerated by orders of magnitude, thus making it feasible for real cases.

This work is carried out under the flat$w$CDM scenario, with the Hubble constant as the only free cosmological parameter. However, we note that the time delays and lens model are also weakly dependent on the other cosmological parameters, such as the matter density of the Universe $\Omega_{\rm m}$, the dark energy equation-of-state parameter $w$, etc. Therefore, glSNe are expected to put constraints on different cosmological parameters simultaneously, works can be found in e.g., \cite{grillo24, birrer24}.

\section*{Acknowledgments}

We thank the anonymous referee who provided useful suggestions that improved this manuscript. We acknowledge the support by National Key R\&D Program of China No. 2022YFF0503403, the Ministry of Science and Technology of China (grant Nos. 2020SKA0110100), and the support from NSFC of China under grant 12533008. This work is supported by China Manned Space Project (No. CMS-CSST-2021-A01, CMS-CSST-2021-A04, CMS-CSST-2021-A07, CMS-CSST-2023-A03). HYS acknowledges the support from NSFC of China under grant 11973070, Key Research Program of Frontier Sciences, CAS, Grant No. ZDBS-LY-7013 and Program of Shanghai Academic/Technology Research Leader. RL acknowledges the support of National Nature Science Foundation of China (Nos 11988101,12022306), CAS Project for Young Scientists in Basic Research (No. YSBR-062). This work was supported by CNES, focused on HST. This work utilizes gravitational lensing models produced by Natarajan \& Kneib (CATS). This lens modeling was partially funded by the HST Frontier Fields program conducted by STScI. STScI is operated by the Association of Universities for Research in Astronomy, Inc. under NASA contract NAS 5-26555. The lens models were obtained from the Mikulski Archive for Space Telescopes (MAST).

\section*{Data Availability}
The data underlying this article will be shared on a reasonable request to the authors.









\bibliography{main}
\bsp	
\label{lastpage}
\end{document}